\documentclass{article}
\pdfoutput=1
\usepackage{dcase2021_techrep,amsmath,graphicx,url,times,booktabs, tabularx}

\title{Bioacoustic Event Detection with prototypical networks and data augmentation}

\name{Mark Anderson, Naomi Harte}
\address{SIGMEDIA\\
         School of Engineering, Trinity College Dublin\\
         Dublin 2, Ireland\\
         anderm3, nharte [@tcd.ie] }

\begin{document}

\ninept
\maketitle

\begin{sloppy}

\begin{abstract}
This report presents deep learning and data augmentation techniques used by a system entered into the Few-Shot Bioacoustic Event Detection for the DCASE2021 Challenge. The remit was to develop a few-shot learning system for animal (mammal and bird) vocalisations. Participants were tasked with developing a method that can extract information from five exemplar vocalisations, or shots,  of mammals or birds and detect and classify sounds in field recordings.  In the system described in this report, prototypical networks are used to learn a metric space, from which classification is performed by computing the distance of a query point to class prototypes, classifying based on shortest distance. We describe the architecture of this network, feature extraction methods, and data augmentation performed on the given dataset and compare our work to the challenge's baseline networks.
\end{abstract}


\section{Introduction}
\label{sec:intro}
Few-shot learning has emerged as a promising tool for sound event detection, and is particularly relevant for bioacoustics applications. Few-shot learning systems must generalise to new classes unseen during training, given only a few labelled instances of each class. This makes them highly suitable to tasks such as the monitoring of animal populations through their vocalisations, where the data for a dedicated classifier may be costly to annotate, and may lose some generalisation.

In the DCASE 2021 Few-shot Bioacoustic Event Detection challenge, participants are asked to provide the onset and offset times of events in a given audio file, given 5 labelled examples of the class of interest. Four separate data sets are provided for training, with a further two for validation. The training set comprises of 14 hours, 20 minutes and contains 19 separate classes, and 4686 events. These classes include bird species, hyenas, and meerkats. The validation set contains 5 hours of audio, 4 classes and 310 events. There is no overlap between these two development sets. Further details are provided in \cite{Challenge} and \cite{challengeData}. Evaluation Data contains only the first 5 positive events for each file, with the aim to use the system to predict all other positive events \cite{challengeEval}.

Prototypical networks can be applied to a few-shot learning problem such as the task proposed in the challenge. They are considered a model based approach to the problem of few-shot learning by Wang et al \cite{wang2020generalizing}, which aims to reduce to constrain the hypothesis space using prior knowledge. This is achieved by learning a non-linear mapping between inputs and a metric/embedding space. Class support points should then be sufficiently separated from each other and a prototype representation of the classes constructed \cite{prtotypicalPaper}. This prototype is constructed as the mean of the support set (which are the labelled instances of the new unseen data to be classified). Classification of each query point is done by finding its nearest class prototype according to some distance function $\mathit{d_\phi}(\mathbf{x}, \mathbf{x'})$, typically the distance function employed is Euclidean distance. 

In this regard, prototypical networks are similar to clustering algorithms and nearest neighbour classification in particular, where the prototype representations of each class are the neighbours to the query point. The efficiency and simplicity in these networks makes them an appealing approach to the problem posed in this challenge.

\section{Implementation Details}
\label{sec:implement}
As mentioned in both Section \ref{sec:intro} and the challenge description, few-shot learning is a novel task. Several methods have emerged as solutions to the task, such as matching networks, however we felt prototypical networks offered the most promising approach to the problem. As such, we propose in this section a modified version of the baseline Prototypical Network, implementing a modified network architecture, as well as data augmentation to supplement training data and validation data.

\subsection{Data Preparation}
All audio files in both the development and evaluation sets are first resampled to a sampling rate of 22050Hz and normalised prior to their transformation into Mel spectrograms. We do not apply any band pass filtering to the audio signals due to the variety of species in the available datasets. Future work could apply a custom filter to each dataset based on the typical frequency ranges present, but as this network is designed to generalise to new, unseen classes, we decided against implementing this for this submission. Each audio file is transformed to a Mel spectrogram, with 128 Mel bins, an FFT size of 1024 samples and a hop of 256 samples. This is the same configuration present in the baseline system.

We also preform Per-Channel Energy Normalisation (PCEN) on the spectrograms (proposed as method of improving robustness to channel distortion in keyword spotting tasks \cite{wang2017trainable}), in order to both reduce noise in the spectrogram and provide normalisation, gain control and compression the file. PCEN has been proposed as an alternative logarithmic scaling of the Mel spectrograms \cite{pcenSEDbio}. PCEN does reduce the noise present in the spectrogram, but non-stationary sources of noise still pose an issue and can contaminate features. In our experiments, we tested our systems using both log-Mel spectrograms and PCEN Mel spectrograms.

\subsection{Data Augmentation}

\begin{figure*}[ht]
\centering
\includegraphics[width=\textwidth]{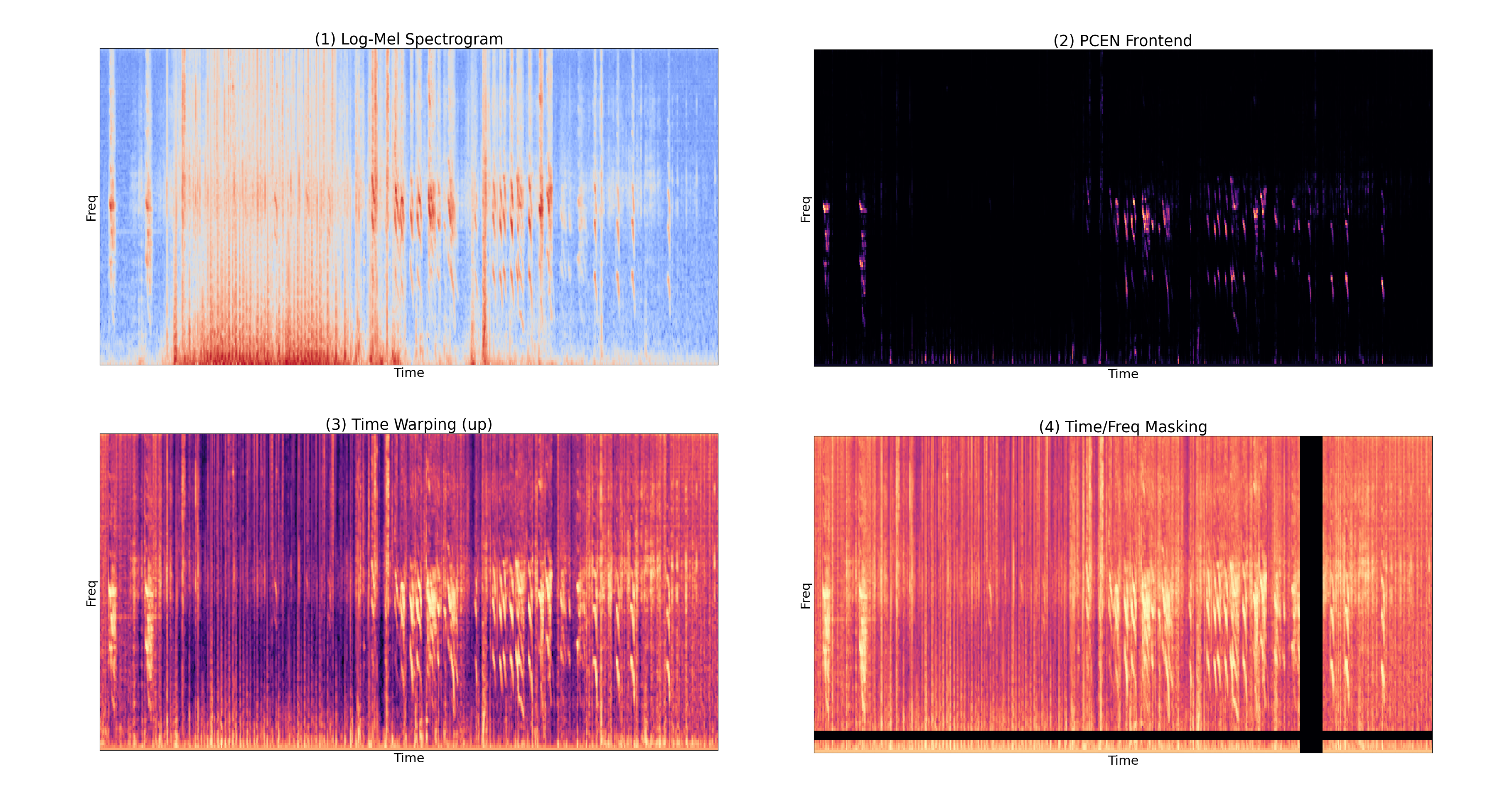}
\caption{\label{fig:aug}Example spectrograms on a 10s clip from the JD subset, (1) Log-Mel Spectrogram, (2) PCEN Front-end, (3) Time warping by 20\%, (4) Mixture of time and frequency masking. PCEN representations when converted to an image, are more difficult to see than log scaled spectrograms, therefore we have decided to present the augmentations using log-scaled spectrograms.}
\end{figure*}

To increase the model performance and create an embedding space that can generalise to loss of information, temporal/frequency shift and new classes, we apply various data augmentation techniques to the resulting spectrogram representations. These data augmentations are based upon work in \cite{specAugment} and \cite{DCASELasseck}. These augmentations are achieved using pyTorch's Audio toolset, Torchaudio, which uses the methods specified in \cite{specAugment}.

\begin{itemize}
    \item \textbf{Time stretching} - We apply a time stretch to each spectrogram, both shortening and lengthening of the spectrogram without any change in pitch info. This translates to a warping along the horizontal axis of the spectrogram according to a specified factor. In our system we have chosen a stretching factor of 5\%, corresponding to two augmented spectrograms at effectively 0.95x and 1.05x the playback speed of the original audio file.
    \item \textbf{Time masking} - Time masking is applied such that $T$ consecutive time steps are masked, with all frequency information set to 0 within that interval. The interval is randomly chosen from a uniform distribution, as per Torchaudios implementation. Due to the considerable length of each file in the data set, this is applied to each 10s chunk of audio, before being reconstructed prior to dataset construction.
    \item \textbf{Frequency masking} - Frequency masking performs a similar augmentation on the vertical, or frequency axis of the spectrogram. Again this is applied in chunks of 10s so as to increase the variability in augmented data.
\end{itemize}

Examples of these augmentations, and the PCEN front-end, are seen in Figure \ref{fig:aug}. Note, that while we show both time masking and frequency masking in the one spectrogram to save space, we did not perform this augmentation on the data itself. 

These augmentations create more data to train and validate the model during development. After augmentation, the validation and evaluation data is transformed into the appropriate dataset form, comprising of positive data points to construct the positive class prototype, negative data points to construct the negative class prototype, and all remaining data to be query points. To prevent class imbalance when training, Random Over Sampling is employed. This is necessary as there is a severe skew in the class distribution of the data, with the majority of occurrences belonging to the negative class. 

\subsection{Model Architecture}
The submitted model architecture is similar to that proposed in the baseline, we have modified the encoder to a less deep model, only using 3 ConvBlocks in place of the 4 in the baseline. Each layer grouping is made up of a Conv2D layer, Batch Normalisation, a ReLU activation layer, and followed by a 2D Max Pooling layer, with pool-size (2, 2). Table \ref{tab:architecture} details the resulting architecture and the ConvBlock sub-block.

\begin{table}[]
\caption{\label{tab:architecture}Architecture of ConvBlock layer and Encoder Architecture\\[0.5em]}
\centering
\begin{tabular}{@{}|l|cc|@{}}
\toprule
\multicolumn{3}{|c|}{\textbf{ConvBlock Architecture}}              \\ \midrule
\textbf{Sub-Layer 1}      & \multicolumn{2}{c|}{Conv2D}           \\
\textbf{Sub-Layer 2}      & \multicolumn{2}{c|}{BatchNorm}        \\
\textbf{Sub-Layer 3}      & \multicolumn{2}{c|}{ReLU}             \\
\textbf{Sub-Layer 4}      & \multicolumn{2}{c|}{MaxPool2D((2,2))} \\ \midrule
\multicolumn{3}{|c|}{\textbf{Encoder Architecture}}                \\ \midrule
\textbf{Layer 1}          & ConvBlock             & 128          \\
\textbf{Layer 2}          & ConvBlock             & 128          \\
\textbf{Layer 3}          & ConvBlock             & 128          \\ 
\textbf{Layer 4}          & Flatten               &              \\ \bottomrule
\end{tabular}
\end{table}

\subsection{Training}
Training of the prototypical network involves minimisation of the negative log-probability of a point to its true class, achieved using the Stochastic Gradient Descent (SGD) algorithm with an initial learning rate of $0.01$ and a momentum factor of $0.85$. Learning rate is scheduled to halve when a plateau is reached, with a patience of 5 epochs and a threshold of 0.01. In our experiments this provided the best environment for training. All our models are trained using 150 epochs, however the best performing model had minimum loss at epoch 126. As we are using additional augmented data, training time can be increased without risk of over-fitting to the training data. 

Training is performed in episodic batches using a custom episodic batch sampling function, easily added to the pyTorch framework. Each episode is randomly generated from the available data, creating a randomised support and query set. Prototypes are computed from the support samples and loss on query points is calculated as per Equation \ref{eqn:loss}.

\begin{align}
    \label{eqn:loss}
    J \leftarrow J + &\frac{1}{N_{C}N_{Q}} \big[\mathit{d}(\mathit{f_{\phi}}(\mathbf{x}), \mathbf{c}_k) + \log \sum_{k'} e^{-\mathit{d}(\mathit{f_{\phi}}(\mathbf{x}), \mathbf{c}_{k'})} \big] \\
    \text{Where, } \nonumber\\
    &\mathit{d}(\cdot) \rightarrow \text{ Euclidean distance function} \nonumber \\
    &\mathit{f_{\phi}} \rightarrow \text{ embedding of query set} \nonumber\\
    &\mathbf{c}_k \rightarrow \text{ class prototype for class } k \nonumber\\
    &N_{C}, N_{Q} \rightarrow \text{ number of classes and query points}\nonumber
\end{align}

In Figure \ref{fig:loss} we have provided the average validation loss during training of a model trained on Mel spectrograms using log-scaling and no augmentation, Mel spectrograms using log-scaling and data augmentation, and finally Mel spectrograms using PCEN and data augmentation. Models using augmented features take longer to train than those which do not, and the usage of PCEN further increases the time necessary for the model to converge on a minimum. Although models using augmented data and PCEN do not achieve lower loss than the original data using log-Mel spectrograms, as seen in Section \ref{sec:results} this does not lead to worse results, in fact they do perform better on this task.

\begin{figure}[t]
\centering
\includegraphics[trim=30 0 0 0, clip=true, width=0.475\textwidth]{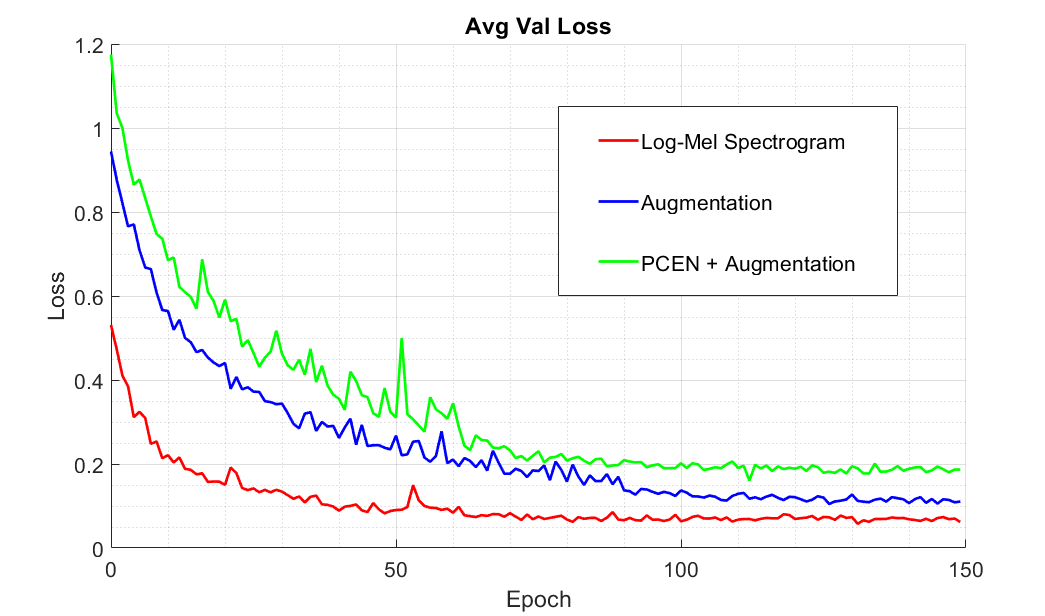}
\caption{\label{fig:loss}Average validation loss for system with log-Mel spectrograms, augmentation, and PCEN + augmentation.}
\end{figure}

\subsection{Evaluation \& Post-Processing}
Evaluation of the model involves calculating the probability of whether a query point belongs to the positive class, achieved by taking the softmax of the distances between query points and the two class prototypes as in Equation \ref{eqn:softmax}.  

\begin{align}
    \label{eqn:softmax}
    \mathit{p_{i}}(y = c_{k} | \mathbf{x}) &= \frac{\exp(-\mathit{d}(\mathit{f_{\phi}}(\mathbf{x}), c_k))}{\sum_{k'}\exp(-\mathit{d}(\mathit{f_{\phi}}(\mathbf{x}), c_{k'}))}
\end{align}

$\mathit{\mathbf{p}}$ is the set of probabilities for each iteration (Equation \ref{eqn:set}). Five iterations of this operation are performed, after which the average of the probabilities is taken. The chosen class is the one with the highest mean probability over all iterations as per Equation \ref{eqn:expected}. Thresholding of probabilities is set to 0.5, although this could be tuned depending on application, and whether recall or precision is valued more.

\begin{align}
    \label{eqn:set}
    &\mathit{\mathbf{p}} = \{\mathit{p_{1}}, ..., \mathit{p_{i}}\}\\
    \label{eqn:expected}
    &\mathit{P_{\phi}}(y = c_{k} | \mathbf{x}) = E \big[\mathit{\mathbf{p}}\big]
\end{align}

For a given file, this results in a classification of each segment as either positive or negative. To prevent abrupt transitions where the classifier may have incorrectly classified a segment in the middle of a vocalisation, median filtering with a window length of 5 samples is applied. Edge detection is performed using a 1D edge detection kernel, convoluted with the classification results at the output of the median filtering stage (Equations \ref{eqn:edge} and \ref{eqn:conv}) to determine the points where onset the onset and offset of an event occurred. The output of this convolution contains the value $1$ at the index of onset, and $-1$ at the index of offset.

\begin{align}
    \label{eqn:edge}
    &k[n] = [1, -1] \text{, edge detection kernel} \\
    \label{eqn:conv}
    &(y \ast k)[n] = y[n] - y[n - 1]
\end{align}

Onset and offset times may then be calculated and written to a CSV file. These onset and offset times are further post-processed using a method proposed by the baseline, where any positive instances that are shorter than 60\% the length of the shortest positive labelled instance, are dropped from the final results. Evaluation of results with and without this post-processing step show that it reduces the amount of false positive events with little change in the amount of true positives registered. Results of experimentation with the minimum allowed event length vary by file, and may warrant future investigation.

\section{Results}
\label{sec:results}
Performance is evaluated on the validation data, provided in the development set released as part of the challenge. We utilise the evaluation method provided by the challenge organisers, which creates a confusion matrix for each file, and calculates precision, recall, and the F-measure.

Given the similarity of our system to that of the baseline, we decided to reproduce the baseline prototypical methods results for comparison with ours. Unfortunately, we could not replicate the results stated in the challenge, achieving at most an F-measure of  $30.437\%$.

In Table \ref{tab:results}, we summarise the results on the validation data provided in the development set on the baseline, and three progressive iterations of our system (Log-Mel spectrogram, Log-Mel Spectrogram with augmentation, Mel spectrogram with PCEN and augmentation).

The best performing model utilises both PCEN on the front-end, and data augmentation to give our best result of $26.243\%$ on the F-measure metric. Breakdown of results by each subset provided for validation can be seen in Table \ref{tab:sub}.

\begin{table}[t]
\centering
\caption{\label{tab:results}Table of results on validation set.}
\begin{tabular}{c|c|c|c}
\hline
\textbf{Model}                    & \textbf{F-Meas. (\%)} & \textbf{Pre.} & \textbf{Rec.}  \\ \hline
\textbf{Baseline Reproduction}    & 30.437                  &   0.456              &  0.228              \\
\textbf{Log Mel Spectrograms}     & 16.565                  & 0.108              & 0.355            \\
\textbf{Data Aug.}        & 20.517                  & 0.143              & 0.360            \\
\textbf{Data Aug. + PCEN} & 26.243                  & 0.200              & 0.381            \\ 
\end{tabular}
\end{table}

\begin{table}[t]
\centering
\caption{\label{tab:sub}Results by validation data subset.}
\begin{tabular}{c|c|c|c}
\hline
\textbf{Data Sub-Set} & \textbf{F-Meas. (\%)} & \textbf{Pre.} & \textbf{Rec.} \\ \hline
\textbf{PB}           & 22.222                  & 0.201              & 0.249           \\
\textbf{HV}           & 32.039                  & 0.199              & 0.825          
\end{tabular}
\end{table}

\section{Discussion}
\label{sec:discuss}
The model architecture we have chosen is similar to that of the baseline, however we have applied various data augmentation methods to the training data to increase model robustness. 

Our aim of improving on the baseline results has not been met, as our F-measure is $\approx4.2\%$ below that of our reproduced baseline. We have made significant improvements to recall, our system is capable of recognising $38\%$ of calls in the validation set, but this has come at the cost of precision, as we have many false positives in our predictions. Future work to improve this may focus on the post processing applied to the data after initial inference. We may also employ deeper models in the future, as there is the possibility that the learned embedding space does not separate the two classes well enough, with suppor points not being sufficiently far away from each other.

Improvement was mostly gained through the usage of PCEN over the more traditional approach of log-scaling the Mel spectrogram, lending evidence to the claim in \cite{pcenSEDbio} that PCEN can significantly outperform log-Mel scaling when used on noisy data, with no significant increase in computational complexity. Our methods of data augmentation also improved on our initial log-Mel spectrogram model, and was further improved by the addition of PCEN. Other methods of data augmentation (such as additional noise from other files, piece-wise frequency stretching, etc.), and augmentation of support points for evaluation should also be investigated. Interestingly, although the average validation loss of models trained using data augmentation was greater than that of the log-Mel spectrograms, their performance when evaluating the validation set was greater. We also found that they could be trained for longer, reaching a minimum less quickly than the original log-based features without augmentation or PCEN.

To conclude, our results of our system show that few-shot learning for bioacoustics applications with prototypical networks is still a novel and challenging task which is in its infancy, but one which can rapidly improve going forward.

\section{Acknowledgements}
\label{sec:ack}
The authors would like to thank the organisers of this challenge, for proposing an interesting and novel problem which is relevant to our own research. We also want to thank the ADAPT Centre for the usage of their 'Boole' cluster when developing this system. The ADAPT Centre for Digital Content Technology is funded under the SFI Research Centres Programme (Grant 13/RC/2106) and is
co-funded under the European Regional Development Fund.

\newpage
\bibliographystyle{IEEEtran}
\bibliography{refs}

% -------------------------------------------------------------------------
% Either list references using the bibliography style file IEEEtran.bst
% \bibliographystyle{IEEEtran}
% \bibliography{refs}
% %
% % or list them by yourself
% % \begin{thebibliography}{9}
% % 
% % \bibitem{dcase2016web}
% %   \url{http://www.cs.tut.fi/sgn/arg/dcase2016/}.
% %
% % \bibitem{IEEEPDFSpec}
% %   {PDF} specification for {IEEE} {X}plore$^{\textregistered}$,
% %   \url{http://www.ieee.org/portal/cms_docs/pubs/confstandards/pdfs/IEEE-PDF-SpecV401.pdf}.
% %
% % \bibitem{PDFOpenSourceTools}
% %   Creating high resolution {PDF} files for book production with 
% %   open source tools, 
% %   \url{http://www.grassbook.org/neteler/highres_pdf.html}.
% %
% % \bibitem{eWilliams1999}
% % E. Williams, \emph{Fourier Acoustics: Sound Radiation and Nearfield Acoustic
% %   Holography}. London, UK: Academic Press, 1999.
% % 
% % \bibitem{ieeecopyright}
% %   \url{http://www.ieee.org/web/publications/rights/copyrightmain.html}.
% %
% % \bibitem{cJones2003}
% % C. Jones, A. Smith, and E. Roberts, ``A sample paper in conference
% %   proceedings,'' in \emph{Proc. IEEE ICASSP}, vol. II, 2003, pp. 803--806.
% % 
% % \bibitem{aSmith2000}
% % A. Smith, C. Jones, and E. Roberts, ``A sample paper in journals,'' 
% %   \emph{IEEE Trans. Signal Process.}, vol. 62, pp. 291--294, Jan. 2000.
% % 
% % \end{thebibliography}

\end{sloppy}
\end{document}